\documentclass[letter,traditabstract]{aa} 
\usepackage{graphicx}

\usepackage{natbib}
\bibpunct{(}{)}{;}{a}{}{,}
\usepackage{txfonts}
\begin{document}
   \title {First starlight spectrum captured using an\\integrated photonic micro-spectrograph}

   \author {N. Cvetojevic,
          \inst{1}\fnmsep \inst{2}\fnmsep \inst{3}\fnmsep \thanks{nick.cvetojevic@mq.edu.au}
          N. Jovanovic, 
          \inst{1}\fnmsep \inst{2}\fnmsep \inst{4}
          C. Betters,
          \inst{5}\fnmsep \inst{6}
          J. S. Lawrence,
          \inst{1}\fnmsep \inst{2}\fnmsep \inst{4}
          S. C. Ellis,
          \inst{4}
          G. Robertson,
          \inst{5}\fnmsep \inst{6} \\      
          \and
          J. Bland-Hawthorn \inst{5}\fnmsep \inst{6}
          }

   \institute{MQ Photonics Research Centre, Department of Physics and Astronomy, Macquarie University, NSW 2109, Australia
         \and
             Centre for Astronomy, Astrophysics and Astrophotonics, Macquarie University, Australia
         \and
             Centre for Ultrahigh Bandwidth Devices for Optical Systems (CUDOS)
         \and
         	 Australian Astronomical Observatory (AAO), PO Box 296, Epping, NSW 2121, Australia
         \and
             Sydney Institute for Astronomy (SIfA), School of Physics, University of Sydney, NSW 2006, Australia
         \and
             Institute of Photonics and Optical Science (IPOS), School of Physics, University of Sydney, Australia
             }

   \date{Received}

 
  \abstract
   {Photonic technologies have received growing consideration for incorporation into next-generation astronomical instrumentation, owing to their miniature footprint and inherent robustness. In this paper we present results from the first on-telescope demonstration of a miniature photonic spectrograph for astronomy, by obtaining spectra spanning the entire H-band from several stellar targets. The prototype was tested on the 3.9~m Anglo-Australian telescope. In particular, we present a spectrum of the variable star $\pi$ 1 Gru, with observed CO molecular absorption bands, at a resolving power $R = 2500$ at 1600 nm. Furthermore, we successfully demonstrate the simultaneous acquisition of multiple spectra with a single spectrograph chip by using multiple fibre inputs. }
   {} {} {} {}


\titlerunning{First starlight spectrum captured using an integrated photonic micro-spectrograph}
\authorrunning{N. Cvetojevic et al.}               
               
   \maketitle
%

\section{Introduction}

An integrated photonic spectrograph (IPS) is a miniaturised, monolithic dispersive device. These wafer-based components are typically only several centimetres in size, which makes them robust against misalignments due to environmental factors. Such properties are highly sought after for use in next-generation astronomical instrumentation, because spectrographs for seeing-limited telescopes, for example, are currently based on large, custom-made optics, which are prone to flexure and thermal drift. The possibility of exploiting IPSs in astronomy was proposed as early as 1995 \citep{W95}, but it was not until the technology reached maturity that it was considered to be a viable option \citep{BH2006SPIE}.

Integrated photonic spectrographs are most commonly fabricated via lithographic methods. This essentially involves depositing material onto a substrate through a mask, which is used to outline the required circuit. However, once the mask is created, it is a relatively inexpensive process to mass-produce hundreds or even thousands of miniature spectrographs. This photonic approach to spectrograph design allows for small, mass-fabricated, modular components to be used instead of the large, custom-built elements used in existing spectrographs \citep{ASBH10,PIMMS10}. Furthermore, photonic spectrographs are commonly designed to operate at the diffraction-limit, circumventing the spectrograph-telescope size relation that often plagues conventional spectrographs on large seeing-limited telescopes \citep{BH2006SPIE,ASBH10}.

In 2009, we experimentally demonstrated the feasibility of this technology for astronomy by measuring atmospheric OH emission lines in the H-band (1460-1810 nm) using a single photonic chip, an early IPS prototype \citep{ME09}. More recently, in \citet{ME12}; hereafter CV12, we characterised an improved prototype IPS that could observe multiple sources simultaneously in a laboratory environment, which serves as the basis for the instrument presented in this work. Here we present the first stellar spectral features obtained using a photonic spectrograph from an on-sky test at the 3.9~m Anglo-Australian Telescope (AAT).


\section{Integrated photonic spectrograph and telescope interface}
\label{sec2}

   \begin{figure*}
   \centering
   \includegraphics[width=0.9\textwidth]{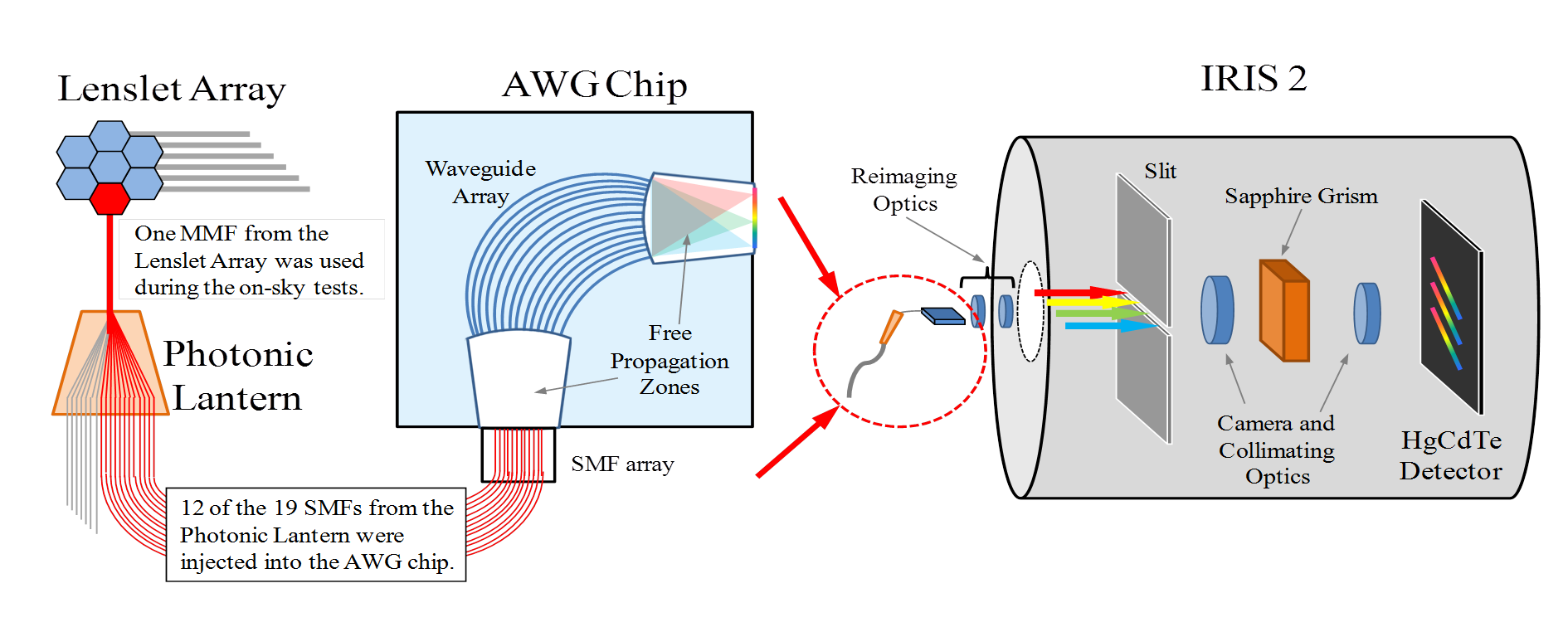}
   \caption
   {Schematic of the IPS interface with the telescope. The array of seven lenslets couples the light from the telescope's Cassegrain focal plane into MMFs. For these on-sky results, only one lenslet, and hence one MMF, was used. The light was transported to a photonic lantern, which converts the multimoded light to a series of single modes in multiple SMFs, which were then simultaneously interfaced to the AWG chip. The AWG chip outputs the spectrally dispersed signal from all 12 input SMFs on its output face, which was re-imaged onto the slit. The IRIS~2 imaging spectrograph was used as a cross-disperser, separating the spectra from the multiple input fibres along with the different grating orders.}
              \label{Fig1}%
    \end{figure*}
The photonic chip we used for telescope tests was a silica-on-silicon wafer, approximately 4~x~4~cm in size with a 3~mm thickness. It consisted of an 8~$\mu$m thick guiding layer embedded in a 40~$\mu$m thick silica cladding, backed by a 2.9~mm thick silicon substrate. The choice of silica for the core/cladding layers was made to maximise transparency at the operating wavelength (H-band). An arrayed waveguide grating (AWG) architecture was used, with the entire photonic circuit confined to the guiding layer. The AWG layout integrates both the collimating and camera optics, and the primary spectral dispersing element into one circuit \citep{OKIMOTO}. Detailed explanations on how the AWG photonic circuitry functions with respect to astronomical requirements can be found in our previous work (\citealt{MESPIE,JONSPIE}; CV12).

Arrayed waveguide grating chips are commonly used in fibre-optic communication systems as wavelength division multiplexors, multiplexing a number of optical carrier signals into a single optical fibre \citep{TAKASHI}. Owing to its roots in the telecommunications industry, the AWG IPS had to be modified to optimise performance for astronomical use (CV12). The input of the AWG chip consisted of a uniformly spaced, 1D array of single-mode optical fibres, which were butt-coupled to the input-free propagation zone (FPZ) of the chip (see Fig \ref{Fig1}.). The light then diffracted within the slab-waveguide of the input FPZ where it was collected by 428 closely spaced single-mode waveguides at the far end. Each of the 428 waveguides in the array were incrementally longer than their neighbour by 27~$\mu$m, such that a path difference was accumulated across the output of the waveguide array. The light was then allowed to interfere in a second slab waveguide FPZ, manifesting a spectrally dispersed signal at the polished output surface of the chip. 
 
Under laboratory conditions, the AWG chip had a resolving power $R = \lambda/\Delta\lambda = 7000 \pm 700$~($0.22 \pm 0.02$~nm spectral resolution) at 1600~nm. The free spectral range (FSR) was 57~nm for the $m = 27^{\textrm{th}}$ diffraction order (CV12), with the spectrum being dispersed over 1.25~mm at the output face, giving a linear dispersion of 45~$\pm$~0.2~nm/mm. Both higher and lower orders were spatially overlaid at the output face of the chip, akin to Echelle gratings that operate at a similarly high order, hence cross-dispersion (secondary wavelength dispersion in a plane perpendicular to the AWG's primary dispersion) was used to separate the orders and obtain a larger overall wavelength coverage. The light was passed through an H band (1460-1810~nm) filter before a sapphire grism (transmission grating bonded to a sapphire prism) was used to cross-disperse the output of the AWG chip. The signal was focused onto an HgCdTe HAWAII-1 CMOS detector using lenses. The detector, filter and grism were part of the IRIS2 imaging spectrograph \citep{IRIS2} located at the AAT.
 
Interfacing the IPS to the AAT was challenging, because direct coupling of the seeing-limited celestial light into a single-mode fibre (used to feed the IPS) at the telescope focal plane was impractical and lossy (as discussed by \citealt{SHAKLAN} and \citealt{CORBETT}). This is primarily because the typical core size of a single-mode fibre (SMF) is an order of magnitude smaller than the size of a seeing-limited point spread function (PSF) at typical focal planes for large astronomical telescopes.This is why seeing-limited telescopes typically use the much larger multimode fibres (MMFs) for efficient light collection at the focal plane. 

Because this is a common predicament faced by most astrophotonic devices, a practical solution using a transitional multi-fibre taper known as a photonic lantern \citep{Noordegraaf09,Noordegraaf10} has been demonstrated \citep{NATCOM}. The photonic lantern acts as an efficient multimode to single-mode converter, allowing for the N-modes of the multimode fibre to be split into N individual single mode cores. Hence, we were able to convert the light fed from a 50~$\mu$μm core MMF into nineteen separate SMFs, twelve of which were spliced (fibres fused together) to the input array of the IPS. A limit of 12 input fibers was placed on the IPS so as to ensure no overlap between neighbouring orders after cross-dispersion, the grounds of which are discussed in some detail in our previous work (CV12). 

The light was injected at the telescope's Cassegrain focal plane into MMFs using a seven-element hexagonal lenslet-array  (see Fig \ref{Fig1}.). Each lenslet sampled 0.4~arcsec of the sky (flat-to-flat), with the entire array sampling $\sim$1.2~arcsec, which is a reasonable match to the median seeing at Siding Spring Observatory of 1.6~arcsec. The MMFs transported the light to a gravitationally invariant platform at the telescope's base that housed the instrument, including the photonic lantern, AWG chip, and  IRIS2.

\subsection{IPS and interface efficiency} 

The telescope interface used to test the IPS on-sky was unoptimised, but was sufficient to obtain a spectrum. In particular, for the results presented in this paper, we used a configuration where a single MMF (and hence one lenslet out of the seven available) was used, as seen in Fig~\ref{Fig1}. This meant that the PSF was undersampled; the lenslet of interest coupling 5-7$\%$ of the PSF flux, and dependent on seeing. 

The on-sky throughput measurements were conducted using a standardised spectral source (Xe lamp) and calibrated using IRIS2. The coupling efficiency of the fibre-optics, which included the lenslet losses, was determined to be $62\pm1\%$. The transmission efficiency of all fibre splices within the optical train as well as the throughput of the photonic lantern and the coupling from the SMF array into the chip was measured to be $14\pm3\%$. The component of the losses due to the lantern took into account the fact that only 12 of the 19 output fibres were used. To account for this, the lantern throughput was simply scaled by 12/19, which was an appropriate approximation because the integration times were significantly longer than the coherence time of the atmosphere. The normalised fibre-to fibre variation in throughput of the SMFs was $30\%$.

We probed the throughput of the AWG chip by coupling a narrow bandwidth laser, with real-time power monitoring, into the chip via a single-mode fibre. A flux-calibrated InGaAs detector was used to image the chip's output, allowing accurate measurements of both the output power and the shape of the point-spread-function, which in turn were used to determine the `native' resolution of the AWG chip (for more details on the setup see CV12).  We measured the total throughput of the AWG-chip itself to be $75\pm5\%$.

The throughput of the cross-disperser (IRIS2) including optics was measured to be $11.6\%$. Therefore, the total throughput of the demonstrator was about $0.03-0.07\%$, with each individual spectra being 1/$12^{th}$ of that ($\sim$0.005$\%$ with up to $30\%$ fibre-to-fibre variation).




\subsection{On-sky IPS performance}

   \begin{figure}
   \centering
   \includegraphics[width=0.4\textwidth]{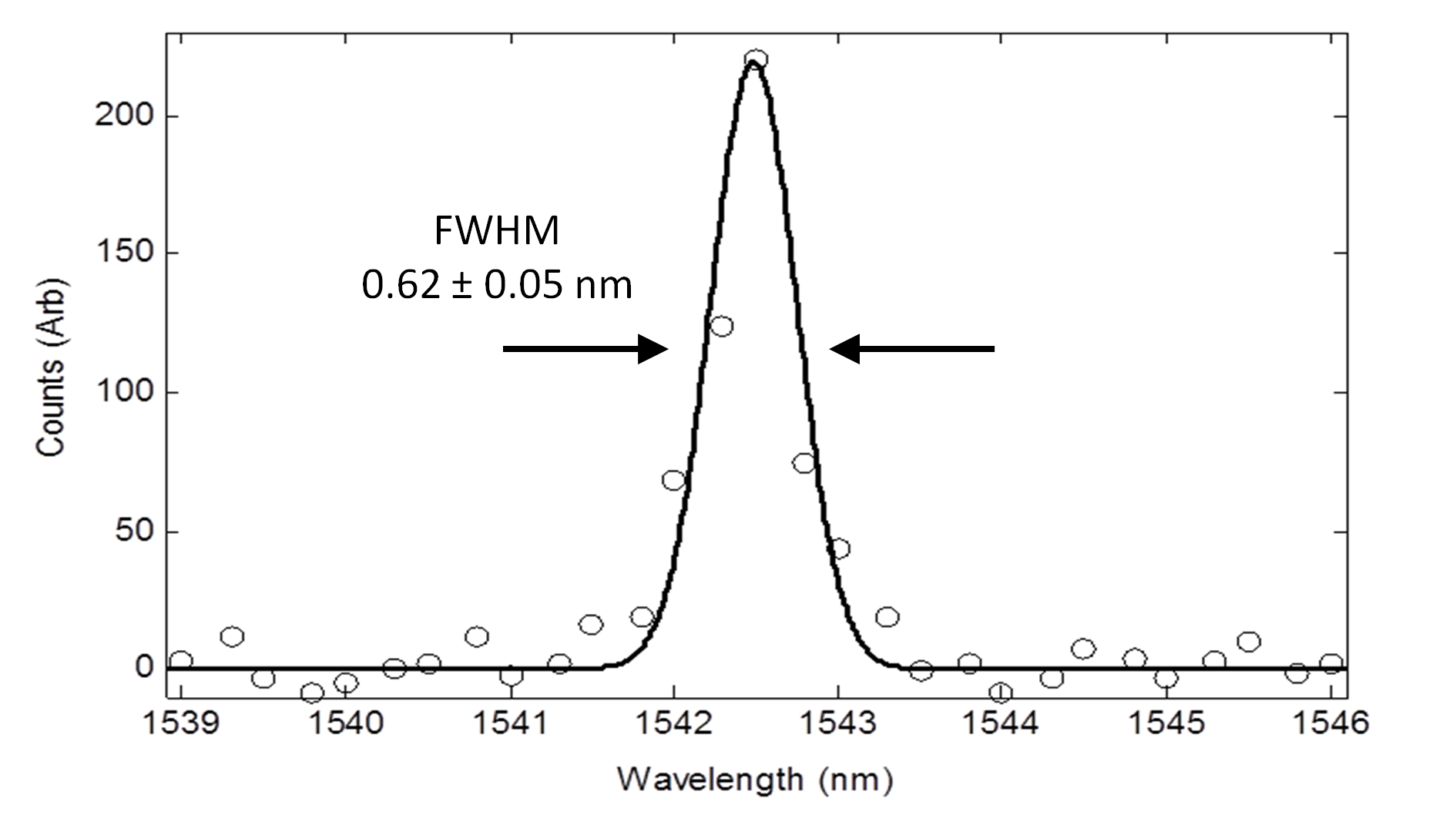}
   \caption{On-sky PSF of the IPS instrument obtained using an Xe arc-lamp emission line. It is not possible to extract detailed information regarding the shape and wings of the PSF from this plot because the PSF was initially undersampled, and artificially broadened by our extraction algorithm, which used a rotation and interpolation function. For details of the raw PSF from the AWG IPS the reader is directed to CV12.}
              \label{Fig2}%
    \end{figure}
While the AWG chip has a native resolving power of 7000~$\pm$~700~(0.22~$\pm$~0.02~nm), when used in tandem with the cross-dispersion optics the on-sky resolving power was reduced to 2500~$\pm$~200 ($\sim$0.6~nm spectral resolution). This was due in part to aberrations from the cross-disperser optics, but more significantly because the pixel pitch of the detector was not matched to, and hence undersampled, the chip's PSF. The resolving power was calculated using a xenon arc-lamp spectrum, taken during the observations for wavelength calibrations, with a typical PSF shown in Fig \ref{Fig2}. 

\section{Results}
\label{sec3}

   \begin{figure}
   \centering
   \includegraphics[width=0.42\textwidth]{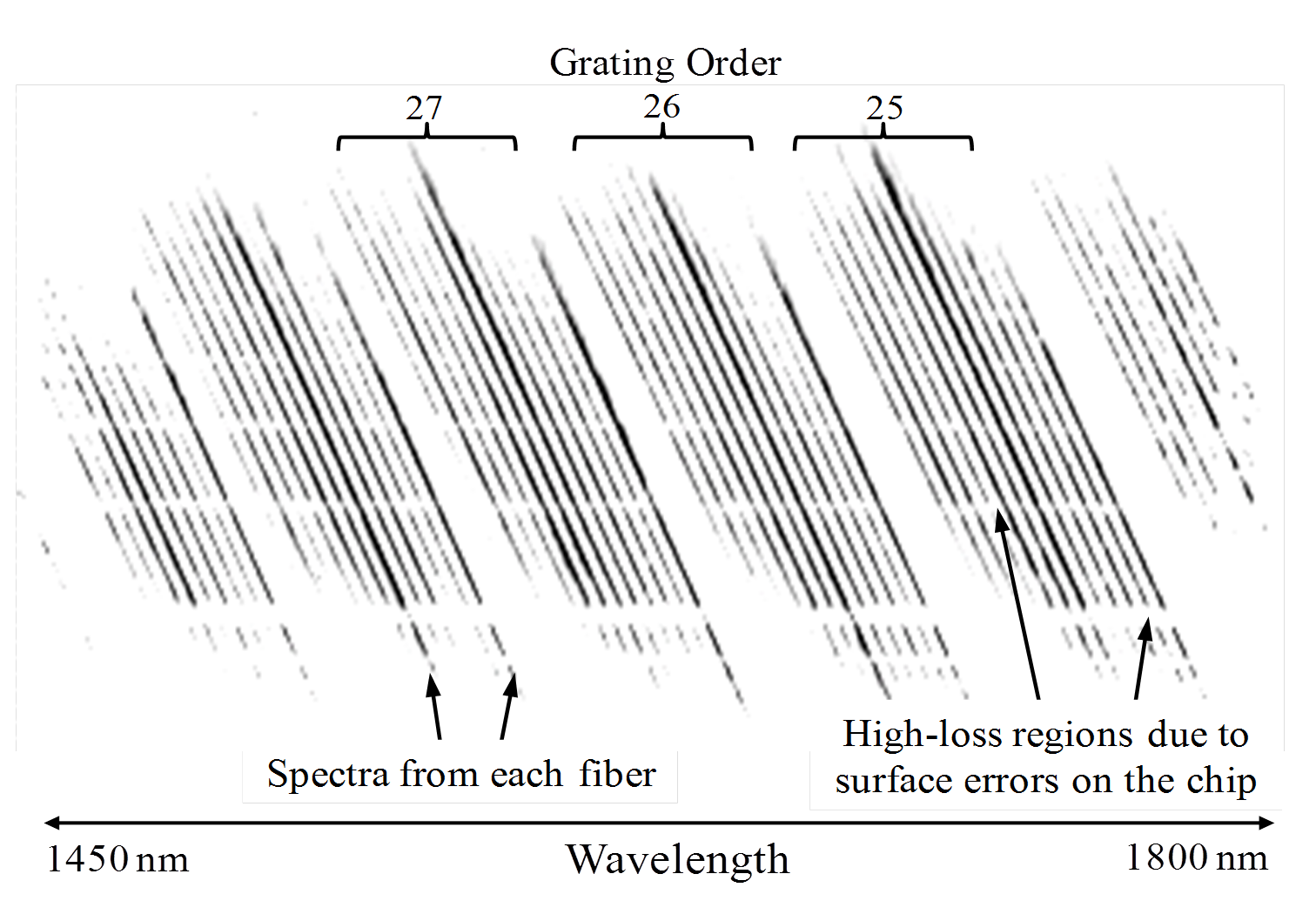}
   \caption{Unreduced readout frame from the IPS of Antares ($\alpha$ Sco), recoloured and contrasted to demonstrate positions of the spectra and the effects of chip imperfections. The AWG dispersion was in the vertical direction, with the cross-dispersion in the horizontal direction. This permits the decoupling of the multiple input signals, as well as the higher AWG diffraction orders. The twelve individual fibre spectra are thin diagonal lines with a small gap separating the higher diffraction orders. The horizontal features running through the frame are caused by imperfections (such as dust and flakes) on the AWG's output facet, which cause scattering. These features were removed by dividing by the instrument flats.}
              \label{Fig3}%
    \end{figure}
%
   \begin{figure*}
   \centering
   \includegraphics[width=0.9\textwidth, height=0.32\textheight]{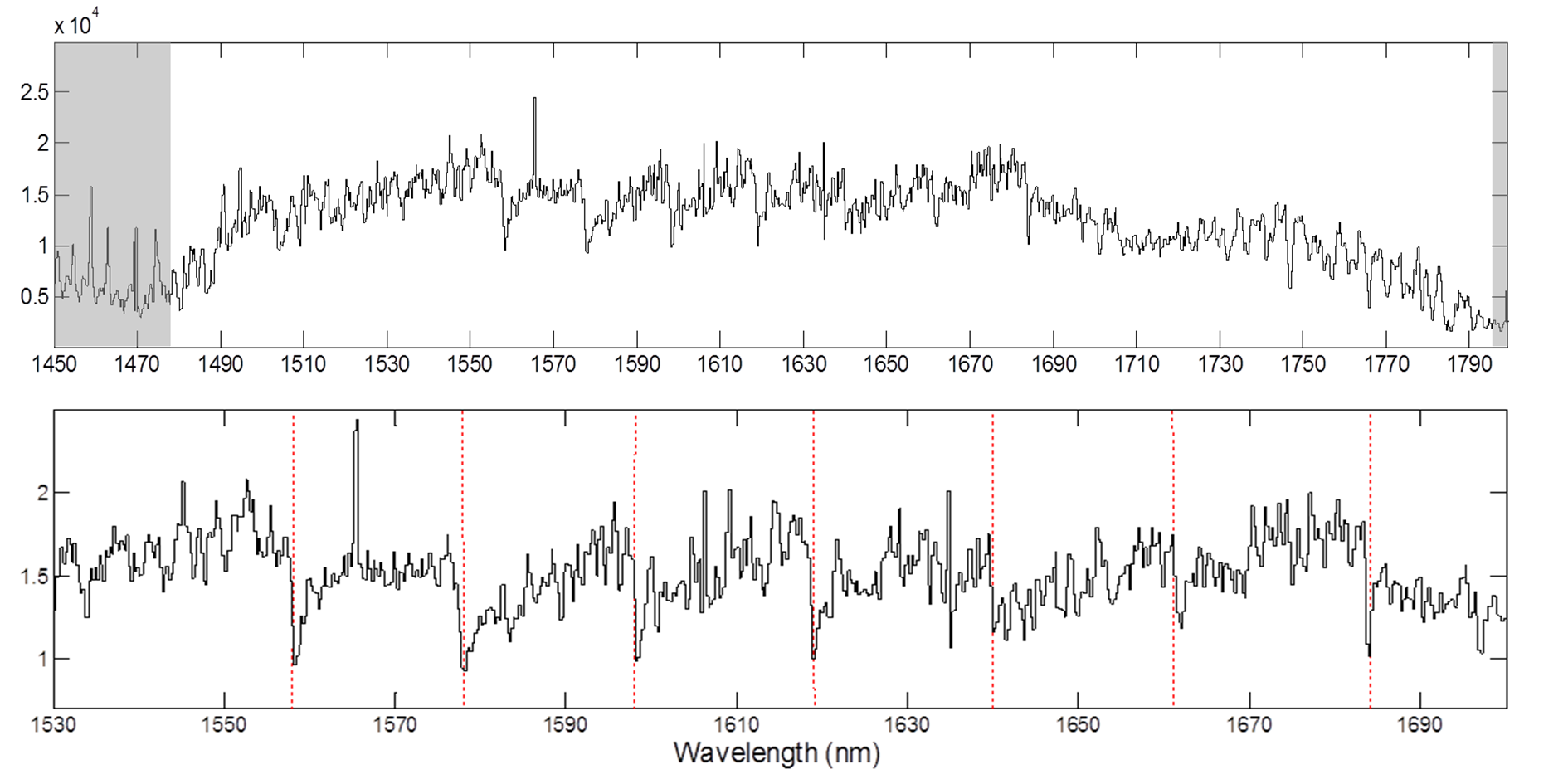}
   \caption
   {H-band spectrum of $\pi$~1~Gru (top) with grey areas indicating the parts of the spectrum contaminated with atmospheric OH absorption, and a cut showing the CO lines in more detail (bottom). The dotted red lines indicate the known wavelength of the CO (3-0) to (9-6) band-heads \citep{OGRILA}.}
              \label{Fig4}%
    \end{figure*}

On-sky testing of the IPS prototype was conducted on 18 May 2011. The unoptimised nature of this demonstrator constrained us to observations of bright stellar sources: $\alpha$~Sco (Antares), $\alpha$~Ara \& $\pi$~1~Gru. Antares and $\alpha$~Ara were primarily used for final on-sky alignment of the telescope interface optics and calibration of the terrestrial atmospheric features, respectively. While we did record their spectra, we found them relatively featureless in the H-band, and consequently do not present them in this paper. 

A typical unreduced readout frame from the IPS is shown in Fig. \ref{Fig3}, where the spectra from each fibre over six diffraction orders can be seen. Because only bright point-sources were observed, all spectra from the fibres are of the same object and are identical; with the exception of a small wavelength shift for the off-centre fibers (CV12). Therefore, it was possible to combine the spectra from each fibre and order into a single spectrum. In total, some 60 individual `mini-spectra' (12 fibres over 5 orders) were extracted from each frame using an automated script, each of which was wavelength-calibrated using Xe lamp emission lines as a reference. Finally they were summed into one complete H-band spectrum. 

The data were additionally reduced by dividing out the instrumental flat-fields that corrected for any instrument response including fibre-to-fibre throughput variations, surface imperfections on the AWG chip, and telescope coupling. We corrected for atmospheric transmission features by normalising to a bright Be star ($\alpha$~Ara), which had no visible features in the H-band.

We observed a type-S star ($\pi$~1~Gru), whose spectrum is shown in Fig. \ref{Fig4}. $\pi$~1~Gru is a variable, late-type star, which is known to contain carbon monoxide (12CO) molecular absorption features in the H-band \citep{DeBeck,OGRILA}. The CO molecular absorption band appears as a sharp absorption feature followed by a decreasing `tail' caused by an unresolved forest of molecular vibrational modes. The results were obtained using an exposure time of 30 minutes. Furthermore, the overall shape of the continuum in the H-band is typical of luminous cool stars, due to H$_{\textrm{2}}$O absorption in the star's atmosphere \citep{LANCON}.
      
These results are to our knowledge the first spectral features from an astronomical source detected using a photonic spectrograph of any kind. The on-sky tests conducted of the IPS demonstrate the potential of the technology in creating miniature and robust photonic spectrographs for astronomical applications.

\section{Conclusions and future applications}

We have demonstrated that it is possible to interface a diffraction-limited photonic spectrograph to a major seeing-limited research telescope and obtain meaningful astronomical spectra. The overall throughput of this early IPS prototype is  poor when compared to modern spectrographs, which was due to the unoptimised nature of the interface optics, but more significantly to the inherent difficulty in coupling a seeing-limited telescope PSF into a photonic spectrograph. Fortunately, these problems could be overcome with a more optimised telescope interface design. Furthermore, because the majority of telescopes in use (or under construction) employ (or plan to employ) an adaptive optics system, a near diffraction-limited PSF will be available, allowing for efficient coupling directly into an SMF, albeit currently over a very small field of view. Hence, a small-field IFU type IPS system could potentially be used without the need for a photonic lantern with throughputs and resolving powers competitive with modern spectrographs, while maintaining the inherent benefits of an all-photonic platform. Space-based applications would obviously be ideally suited to this photonic approach, but remain untested.

If an IPS system could make use of a near diffraction-limited PSF, and operate with a sufficiently broad free-spectral range (100's of nm), it would make cross-dispersion unnecessary. Therefore, the only non-photonic part of our IPS prototype could be removed entirely, and the detector bonded directly to the photonic chip (or multiple stacked chips). This would also require a new generation of detectors with pixel sizes capable of Nyquist-sampling the chip PSF of $\sim$~7~$\mu$m along one axis.

\begin{acknowledgements}
We wish to thank A. Horton, Macquarie Engineering \& Technical Services, and all of the AAO Observatory site staff for support during the observing run. We also wish to thank M. Ireland for helpful insights into H-band spectral features of cold stars. This work was conducted with the assistance of the Australian Research Council (ARC), and the AAO. Elements of this work were supported by the European Union via the Opticon Integrated Infrastructure Initiative of Famework Programme 7. 
\end{acknowledgements}

\bibliographystyle{aa}
\bibliography{Onskypaper}

\end{document}